\documentclass[fleqn,usenatbib]{mnras}

\usepackage{newtxtext,newtxmath}

\usepackage[T1]{fontenc}

\DeclareRobustCommand{\VAN}[3]{#2}
\let\VANthebibliography\thebibliography
\def\thebibliography{\DeclareRobustCommand{\VAN}[3]{##3}\VANthebibliography}

\usepackage{graphicx}	
\usepackage{amsmath}	
\usepackage{amssymb}	

\title[Wide-Binary Origin of the Pluto-Charon System]{The Wide-Binary Origin of The Pluto-Charon System}

\author[Rozner et al.]{
Mor Rozner,$^{1}$\thanks{E-mail: morozner@campus.technion.ac.il}
Evgeni Grishin,$^{1}$
Hagai B. Perets $^{1}$
\\
$^{1}$Technion - Israel Institute of Technology, Haifa, Israel, 3200003
}

\date{Accepted XXX. Received YYY; in original form ZZZ}

\pubyear{2020}

\begin{document}
\label{firstpage}
\pagerange{\pageref{firstpage}--\pageref{lastpage}}
\maketitle

\begin{abstract}
The Pluto-Charon binary system is the best-studied representative of the binary Kuiper-belt population. Its origins are vital to understanding the formation of other Kupier-belt objects (KBO) and binaries, and the evolution of the outer solar-system. The Pluto-Charon system is believed to form following a giant impact between two massive KBOs at relatively low velocities. However, the likelihood of a random direct collision between two of the most massive KBOs is low, and is further constrained by the requirement of a low-velocity collision, making this a potentially fine-tuned scenario. Here we expand our previous studies and suggest that the proto-Pluto-Charon system was formed as a highly inclined wide-binary, which was then driven through secular/quasi-secular evolution into a direct impact. Since wide-binaries are ubiquitous in the Kuiper-belt with many expected to be highly inclined, our scenario is expected to be robust.  We use analytic tools and few-body simulations of the triple Sun-(proto-)Pluto-Charon system to show that a large parameter-space of initial conditions leads to such collisions. The velocity of such an impact is the escape velocity of a bound system, which naturally explains the low-velocity impact. The dynamical evolution and the origins of the Pluto-Charon system could therefore be traced to similar secular origins as those of other binaries and contact-binaries (e.g. Arrokoth), and suggest they play a key role in the evolution of KBOs.
\end{abstract}




\section{Introduction}
The Kuiper-belt hosting numerous Kuiper-belt objects (KBOs) is a relic of ancient era of the Solar system, and henceforth 
preserves valuable clues regarding the dynamics that led the Solar system to its current state. 
KBO-binaries (KBBs) are ubiquitous among KBOs, in particular massive ones, and tens of percents of the current large KBOs are found to be part of bound binary (and satellite) systems \citep{GoldreichLithwickSari,Noll2007,Fraser2017}.

The dwarf-planet Pluto was the first and largest KBO to be discovered \citep{Tombaugh1946}, as well as the first KBB to be found, with the discovery of it most massive companion Charon \citep{ChristyHarrington1978}. 
Recent data from the New Horizons spacecraft sets firmly the size and density of the current Pluto and Charon to be $R_{\rm Pluto}= 1188.3 \pm 1.6 \ \rm{km}, \ R_{Charon}=606 \pm  1\  \rm{km} , \ \rho_{Pluto}=1854 \pm 11\ \rm{kg/ m^3}$ and $\rho_{\rm Charon}= 1701 \pm
33 \ \rm{kg/ m ^3}$ \cite{Nimmo2017}. The mass ratio between Charon and Pluto, which is given by $0.1218:1$, the relatively close distance between them of $\sim 2\times 10^7\ \rm{m}$ \citep{Stern2015_NewHorizon} and the large mutual inclination of $\sim 119^\circ$ to Pluto's orbit \citep{Nao+10} are unique among moons in the solar system. 

There are three major models suggested for the formation of KBBs: gravitational collapse, a giant impact and dynamical capture. The gravitational collapse in-situ formation \citep{Nesvorny2010} requires a large gravitationally unstable pebble cloud -- at least as massive as the Pluto-Charon system -- since mass is lost during the formation. However, simulations showed that the formation of such a particle cloud is unlikely \citep{Johansen2015}.
In the giant impact scenario \citep{Mckinnon1984,McKinnon1989,Canup2005,Canup2011,Desch2015,Sekine2017}, the progenitor of Charon hit proto-Pluto with a velocity comparable to the escape velocity and either merged (immediately/after a rebound \citealp{Leinhardt2012}) with Pluto and ejected a massive disc of debris which then formed Charon, or grazed Pluto and was then directly captured to be the currently observed Charon. In the dynamical capture scenario \cite{GoldreichLithwickSari} two (typically massive) unbound KBOs become bound through a close-passage during which the relative velocities are dissipated through dynamical friction by the planetesimals in their surrounding environment, forming initially very wide-binaries, close to the Hill-radius of the system. Further dissipation could drive the orbit into shorter period.   

The low-velocity collision between unrelated (unbound) and most massive KBOs in the Solar system (proto-Pluto and proto-Charon) is a potentially low-probability event \citep{Canup2005}, given the rarity of such objects,  but it depends on the timing and location (distance from the Sun) of the impact event. Moreover, it is neither likely to explain the formation of the rest of Pluto's moons nor the absence of a fossil bulge \citep{Nimmo2017,McKinnon2017}. Although \cite{KenyonBromley2014} claim that the number of the collisions might be high, one should note that the abundance of objects in the relevant sizes and velocities in the ancient Kuiper-belt is no more than a few tens \citep{Canup2005}. The time of the Pluto-Charon formation is unknown, and is only restricted to occur in the 'pre-installation phase' which took place in the first $500 \ \rm{Myr}$ of the solar-system history \citep{Greenstreet2015}; this uncertainty may give rise to corrections of orders of magnitudes in the estimate presented in 
\cite{Canup2005}. 

In practice, all binaries in the Solar system are a part of a triple system -- with the third companion being the Sun; this provides a fertile ground for significant secular and quasi-secular effects \citep{PeretsNaoz2009,ArokothNature}.  In this view, the Pluto-Charon system is in fact a part of a hierarchical triple, with the Sun as a third distant perturber, which could be important to the dynamics of the Pluto-Charon binary. The significant hierarchy allows to treat the triple system as an inner binary -- the Pluto-Charon binary -- orbited by the outer binary -- the Sun. Here we show that the secular or quasi-secular evolution of the Sun and the proto-Pluto-Charon triple system could naturally give rise to a low-velocity grazing impact at high inclination which is a prerequisite for the formation of the Pluto-Charon system.  It provides an alternative and possibly more robust channel for the origin of this system, consistent with the likely origin of other contact KBBs \citep{PeretsNaoz2009,ArokothNature}. 

We begin with a brief introduction of the role of secular/quasi-secular evolution in the evolution of KBBs (\ref{sec:secular}), followed by an analytical description of the dynamics  of  Pluto-Charon  binary  in  the  different  regimes (\ref{sec:analytic}), we then describe our numerical results (\ref{sec:numerical}), discuss caveats (\ref{sec:caveats}) and summarize (\ref{sec:summary}).

\section{Secular and quasi-secular evolution of KBO binaries}
\label{sec:secular}
The three-body problem is one of the most famous non-integrable problems, tracing back to the pioneering work of Poincare \citep{Poincare1892}.
Fortunately, under certain conditions, some cases could be analysed using perturbative methods. Hierarchical triples are systems that contain an inner binary and a distant tertiary. The system could be described as two binaries -- the inner one and the outer one. When the period of the outer binary is much larger than the period of the inner one, we can average their orbits and consider them as two ellipse shaped mass-wires that interact weakly with each other. This averaging method in the analysis of the secular behaviour is the Lidov-Kozai (LK) mechanism \citep{Lidov1962, Kozai1962}, which gives rise to the exchange of the inner inclination and eccentricity; that could be significant and may lead to extreme orbital evolution, and even flips from prograde to retrograde orbits and vice versa \citep{Naoz2016}. This formalism ignores and averages over timescales shorter than the secular timescale given by \citep{KinoshitaNakai1999,Antognini2015}

\begin{align} \label{tau_sec}
\tau_{\rm{sec}}\approx \frac{8}{15\pi}\frac{m_1+m_2+m_3}{m_{3}} \frac{P_{\rm{out}}^2}{P_{\rm{in}}}\left(1-e_{\rm{out}}^2\right)^{3/2}
\end{align}

where $m_1$ and $m_2$ are the masses of the companions of the inner binary, $m_3$ is the mass of the outer tertiary $P_{\rm{out}}, \ P_{\rm{in}}$ are the periods of the outer and inner binaries correspondingly and $e_{\rm{out}}$ is the eccentricity of the outer binary. 

The standard LK mechanism relies on significantly large separation between the timescales of the inner and outer binaries, $P_{\rm in}/P_{\rm out}$. 
When the systems are mildly hierarchical, short-term effects become important as well, and LK assumptions and results lose accuracy. The evolution is in the quasi-secular regime rather in the 'standard' LK regime \citep{AntoniniPerets2012,Antognini2015,LuoKatz2016,QuasiSecular128}. 

The LK mechanism and its quasi-secular generalization \citep{AntoniniPerets2012} enhance the mergers and collisions of a variety of stellar objects (e.g. see \cite[for a review]{Naoz2016}). In particular, it could play a key role in the  evolution of KBO binaries and formation of short-period and contact-binaries  \citep{PeretsNaoz2009,Nao+10,Por+12,Gri+16,Gri+17,Mic+17,ArokothNature,Lyr+20}. 

The collision velocity that was predicted is comparable to the escape velocity or slightly above \citep{Canup2005} as well as the high inclination of Pluto are suggestive of a possible collision of two-bound objects at high inclination. These are natural outcomes from a secular/quasi-secular evolution and could therefore point to the possible involvement of quasi-secular evolution that drives the formation of the Pluto-Charon binary, similarly to the process that likely formed the contact binary KBO Arrokoth.

The KBO contact binary (2014) $\rm MU_{69}$ (Arrokoth) was discovered by New Horizons search team using the Hubble space telescope, and was chosen as a main target of an extended exploration mission of New Horizons \citep{Stern2018ArokothDiscovery,Stern2019_NewHorizons}.
 One of the leading models of the formation of Arrokoth --- which is supported by new measurements \citep{Stern2019_NewHorizons,Mckinnon2020} -- argues that the formation of Arrokoth arose from a gentle collision between two perturbed wide companions \citep{ArokothNature}. Under certain conditions, wide-binaries could be perturbed significantly such that they could form later contact binaries during secular and quasi-secular evolution. 

In the following we propose and analyze a similar formation channel for the Pluto-Charon system via secular/quasi-secular evolution that leads ultimately to a collision between the inner binary companions -- Pluto and Charon, due to perturbations from the distant perturber -- the Sun.

\section{Analytical description}\label{sec:analytic}
We propose that the current Pluto-Charon short-period binary system originated from a much wider KBO-binary system possibly formed through the gravitational-instability or the dynamical capture scenario, and later evolved through secular or quasi-secular evolution.

Consider an inner wide-binary composed from the progenitors of Pluto and Charon, with separation $a_{\rm in}$, eccentricity $e_{\rm{in}}$ and total mass $m_{\rm in}$, and the Sun as a distant perturber, forming together a hierarchical triple where the separation of the outer binary is given by $a_{\rm out}$ and the mass by $m_{\rm out}$. The dynamical evolution of the system could be described by four main regimes: non-collisional, precession dominated (without collision), collisional secular (or quasi-secular) evolution and collisional non-secular evolution.  
The initial conditions of the system, i.e. initial mutual inclination $i_0$, eccentricity e$_0$ and the separation between the inner binary companions $a_{\rm in}$, determine the regime; or equivalently $j_z=\sqrt{1-e^2}\cos i$ and $a_{\rm in}$, where $i$ is the mutual inclination of the inner binary. Note that from the features of the secular and quasi-secular evolution, $a_{\rm in}$ and $a_{\rm out}$ are approximately conserved through the evolution.  
 We normalize the inner semi-major axis by the Hill radius, $\alpha= a_{\rm{in}}/R_H$. The Hill radius is defined by $R_H=a_{\rm out}(1-e_{\rm out}) \left((m_{\rm{Pluto}}+m_{\rm{Charon}})/3M_{\odot}\right)^{1/3}\approx 6\times 10^6 \  \rm{km}$. 
 
 Hereafter we briefly review the behavior in the different dynamical regimes.

\subsection{Standard LK Oscillations}
When the periods of the inner and outer binaries are well separated, the Hamiltonian of the problem could be decomposed into two Keplerian Hamiltonians and a weak interaction term between the two orbits \citep{Lidov1962,Kozai1962}. Over long timescales, the orbits exchange angular-momentum, while the energy exchange is negligible and the inner and outer semi-major axes remain roughly constant. These conditions induce periodic variation of eccentricity and inclination. 

The motion is governed by the Hamiltonian with the perturbation expanded in multipole expansion \citep{Harrington1968}

\begin{align}
\mathcal H & = \frac{Gm_1 m_2}{2a_{\rm{in}}}+\frac{Gm_3(m_1+m_2)}{2a_{\rm{out}}}+\mathcal H_{\rm{pert}}; \nonumber \\
\mathcal H_{\rm{pert}} & = \frac{G}{a_{\rm{out}}} \sum_{j=2}^\infty \left(\frac{a_{\rm{in}}}{a_{\rm{out}}}\right)^j \left(\frac{r_1}{a_{\rm{in}}}\right)^j \left(\frac{a_{\rm{out}}}{r_2}\right)^{j+1}\mathcal M_{j} P_{j}\left(\cos \Phi\right), \nonumber \\
\mathcal M_j & = m_1 m_2 m_3 \frac{m_1^{j-1}-(-m_2)^{j-1}}{(m_1+m_2)^j}
\end{align}

where $r_i$ is the distance between the two companions of the $i$-th binary, $P_i$ is the $i$-th Legendre polynomial and $\Phi$ is the angle between $r_2$ and $r_1$.

The standard LK formalism considers double-averaging, i.e. averaging over both inner and outer mean anomalies. The averaging is done in  von Zeipel technique  \citep{vonZeipel1916arkiv}, and enables an extraction of the secular changes in the system -- changes in the orbital elements along timescales much longer than the orbital period. The lowest order, and most significant, is the quadruple order. Since the inner binary cannot exceed the Hill radius, the next octupole order is weaker by at least $a_{\rm in}/a_{\rm out} \le (m_{\rm in}/3m_{\rm out})^{1/3}\approx 10^{-3}$, octupole evolution can be safely neglected. We therefore focus only on quadrupole evolution.

In the quadruple order, the maximal eccentricity, and the corresponding minimal inclination, could be written as \citep{Innanen1997}

\begin{align}
e_{\rm{max,LK}}=\sqrt{1-\frac{5}{3}\cos^2 i_0}, \ i_{\rm{min,LK}}= \arccos \left( \pm \sqrt{\frac{3}{5}} \right)
\end{align}

The possible $i_{\rm{min}}$, $39.23^\circ$ and $140.77^\circ$ set the boundaries where LK evolution is active. 

Due to the secular evolution of the distant perturber -- the Sun -- the inner binary, which contains the progenitors of Pluto and Charon, experiences extreme oscillations of its eccentricity and mutual inclination, where the highest eccentricity obtained with the lowest mutual inclination and vice versa. Under the constraints introduced above, the LK mechanism might lead to collisions \citep{PeretsNaoz2009,ArokothNature}.

\subsection{Quasi-secular and non-secular Regime}

The double-averaging approximation breaks down when the system is mildly hierarchical, i.e. when the inner and outer periods become comparable \citep[e.g.][]{AntoniniPerets2012,LuoKatz2016,QuasiSecular128}. The breakdown leads to corrections in the evolution and in particular, corrections of the maximal eccentricity and critical inclinations for onset, which will be indexed by $QS$ \citep{QuasiSecular128}. 

\begin{align}
e_{\rm{max,QS}}= \sqrt{1-\frac{5}{3}\cos^2 i_0 \frac{1+\frac{9}{8}\epsilon_{\rm SA} \cos i_0 }{1-\frac{9}{8}\epsilon_{\rm SA} \cos i_0}}, \\
i_{\rm{min, QS}}=\arccos \left(\pm\sqrt{ \frac{3}{5}}-\frac{27}{40}\epsilon_{\rm{SA}}\right)
\end{align}

were $\epsilon_{\rm{SA}}$ is the strength of the single averaging, given by  eq. \ref{epsilon_SA} \citep{LuoKatz2016,QuasiSecular128}, $\epsilon_{\rm{SA}}=P_{\rm{out}}/2\pi \tau_{\rm{sec}}$ were $P_{\rm{out}}$ is the period of the outer binary and $\tau_{\rm{sec}}$ is defined in eq. \ref{tau_sec}.  

Quasi-secular analysis enables us to treat collisions of less hierarchical systems, i.e. wider inner binaries.

\subsection{Precession due to Oblateness}

Non-spherical shapes of objects lead to corrections in the gravitational potential, which generate an extra precession that might quench the LK mechanism. 
The leading term which encapsulates the dynamics induced by oblateness is the $J_2$ coefficient \citep{MurrayDermottSolarSystem2001}. \cite{Nimmo2017} introduced a detailed study of the observational properties of Pluto and Charon. The measured upper bounds for the oblateness of Pluto and Charon were $0.006$ and $0.005$ correspondingly. For Pluto, the oblateness is proportional to $R_{\rm{Pluto}} \omega^2 /2g$ up to a constant factor of order unity were $R_{\rm{Pluto}}$ is Pluto's radius, $\omega$ is its rotation angular frequency and $g$ is the surface gravity.

\subsection{Roadmap}
\begin{figure*}
    \includegraphics[width=0.99\linewidth]{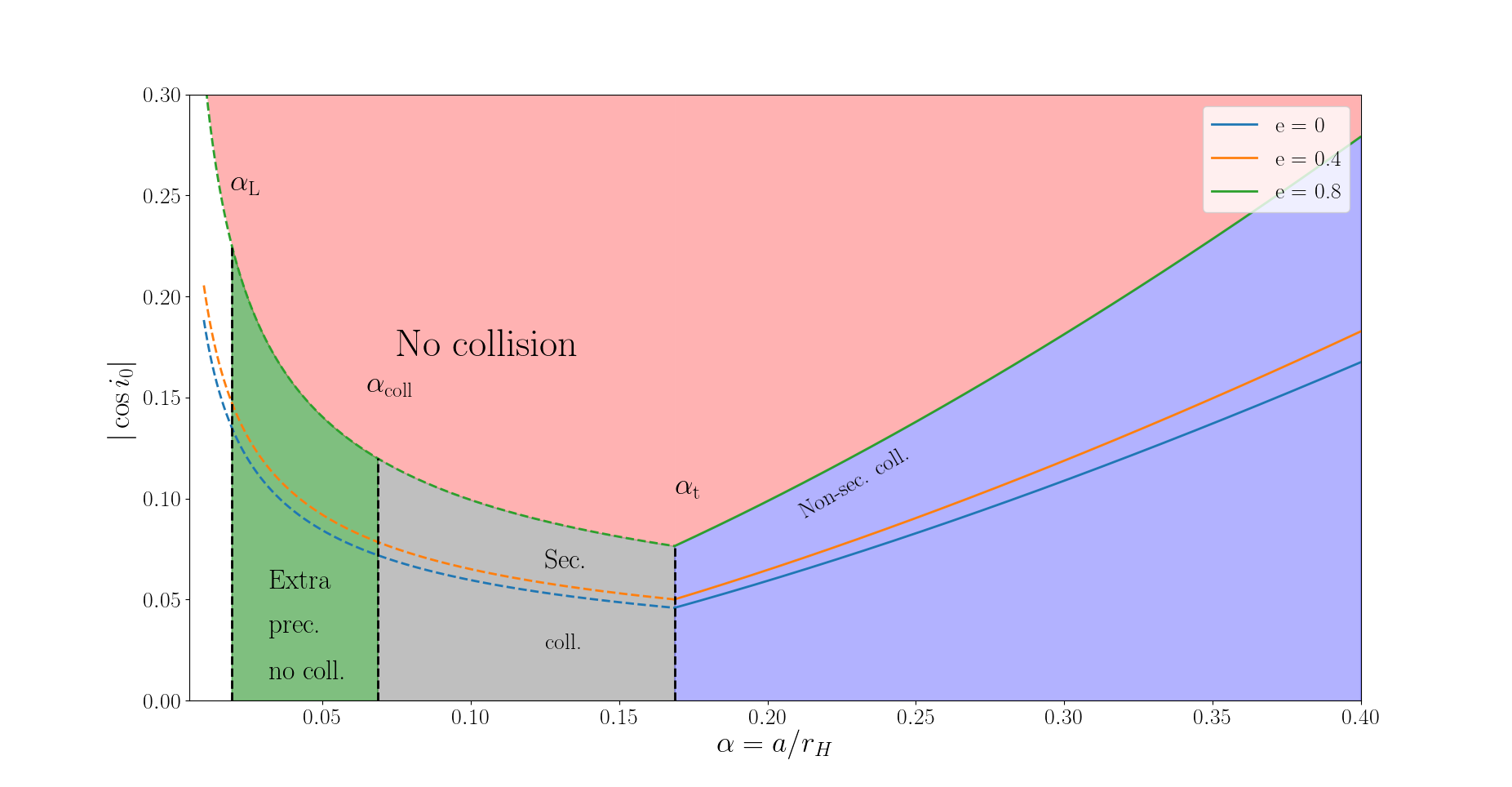}
\caption{Parameter-diagram of the different possible regimes of behaviour of Pluto-Charon system. The colored regimes correspond to the different dynamical regimes of the system: no collision, extra precession no collision, secular and non-secular evolution. The lines correspond to illustrations of systems with different hierarchies -- encapsulated in $\alpha$, and different eccentricities.
} 
\label{fig:phase_diagram}
\end{figure*}

Here we will review the roadmap that describes the transitions between the different regimes, as shown in Figure \ref{fig:phase_diagram}.

Since the objects are not completely spherical, using a point-mass gravitational potential of spherical objects neglects a potentially important aspect of the evolution. Oblateness induces extra apsidal precession on the Keplerian ellipse. When oblateness-induced presession becomes comparable or larger than that induced by the LK evolution (here taken up to quadruple order), the oscillations are quenched and the behaviour is dictated by the precession. For small enough values of $\alpha$, LK oscillations are completely quenched due to oblateness-induced precession. Above a critical value of $\alpha$, signed by $\alpha_L$, LK oscillations become more important, the oblateness-induced precession is non-negligble and prevents the significant eccentricity growth, thereby avoiding collisions; the parameter space corresponds to this scenario is colored with green -- extra precession no collision. The oblateness manifests itself by a dimensionless constant, $J_2$, which is the second Laplace coefficient in the expansion of the potential. 
 The ratio between the LK-induced and oblateness-induced precession is given by \citep{LiuMunozLai2015,ArokothNature}

\begin{align}
\epsilon_{\rm{rot}}= \frac{3}{2}J_2 \frac{m_{\rm{in}}}{m_{\rm{out}}} \frac{a_{\rm{out}}^3(1-e_{\rm{out}}^2)^{3/2}R_{\rm Charon}}{\alpha^5 R_H^5},
\end{align}

 where $J_2=0.005$ is the upper bound of the measured oblateness of Charon \citep{Nimmo2017}.
 Oblateness effects become important when $\epsilon_{\rm{rot}}\gtrsim 1$, and for $\epsilon_{rot}=3/2$ we define Laplace radius as $R_L=\alpha_L R_H$ \citep{Tremaine2009}. 

The quenching of LK oscillations becomes less significant when $\alpha> \alpha_L$, where  $\alpha_L$ is given by 

\begin{align}
\alpha_L=\left(\frac{J_2 m_{\rm{in}} a_{\rm{out}}^3 (1-e_{\rm{out}}^2)^{3/2} R_{tot}^2}{m_{\rm{out}} R_H^5}\right)^{1/5}\approx 
0.02
\end{align}

The transition between the precession dominated regime and the secular-collision regime could be derived from the characteristics of the LK oscillations.
The possible eccentricity regime for a collision to occur is restricted to the range $e_{\rm{min}}=e_{\rm{coll}}$ and $e_{\rm{max}}$; the minimal and maximal eccentricities depend on the geometric configuration of the system, which affect the LK evolution \citep[e.g.][]{KinoshitaNakai1999,PeretsNaoz2009,Naoz2016,Gri+17}

\begin{align}
e_{\rm{coll}}=1-\frac{R_{\rm{tot}}}{\alpha R_{\rm{H}}}, \ e_{\rm{max}}=\sqrt{1-\frac{5}{3}(1-e_{\rm{out}}^2)\cos^2 i_{\rm{out}}}
\end{align}
The lower bound on the eccentricity, enables us to derive a critical inclination, 

\begin{align}
\cos i_0 = \sqrt{\frac{6 R_{\rm{tot}}}{5(1-e_0^2)\alpha R_{\rm{H}}}}
\end{align}

Significant LK oscillations (of an initially circular orbit) occur for inclinations in the range $40^\circ \lesssim i_0 \lesssim 140^\circ$, where large inclinations correspond to small eccentricities and vice versa. 

In the presence of oblate bodies and under the assumption of maximal initial inclination of $\cos i_0=90^{\circ}$, the maximal eccentricity could be estimated by the implicit expression \citep{LiuMunozLai2015,ArokothNature}

\begin{align}
\frac{\epsilon_{\rm rot}}{3}\left(\frac{1}{(1-e_{\rm max}^2)^{3/2}}-1\right)=\frac{9}{8}e_{\rm max}^2,
\end{align}

which could be approximated by $e_{\rm max}\approx 1-\frac{2}{9}\epsilon_{\rm rot}^{3/2}$ under the assumption of large eccentricity ($e_{\rm max}^2\approx 1$) and weak effect of the rotation term ($\epsilon_{\rm rot} \ll 1$). 

Henceforth, the minimal $\alpha$ value for collision (in the secular regime) is given by 

\begin{align}
\alpha_{\rm{coll}}= \left(\frac{2R_H^3  \alpha_L^{10}}{81R_{\rm{tot}}^3}\right)^{1/7}\approx 0.07
\end{align}

The regime which corresponds to 'pure' secular collision is colored with gray -- secular collision.

 The transition between secular and quasi-secular regime can be derived from the strength of the quasi-secular corrections over the strength of the 'standard' LK ones, here taken to quadrupole order. The strength of the perturbations from single-averaging -- averaging over the inner orbit only-- is given by  \citep{LuoKatz2016,QuasiSecular128}
 
 \begin{align}\label{epsilon_SA}
\epsilon_{\rm SA}=\frac{P_{\rm out}}{2\pi \tau_{\rm sec}}=\left(\frac{a_{\rm in}}{a_{\rm out} (1-e_{\rm out}^2)}\right)^{3/2}\frac{M_\odot}{(m_{\rm tot}m_{\rm in})^{1/2}} \approx \\ \nonumber
\approx  \frac{\alpha^{3/2}}{\sqrt{3}(1+e_{\rm out})^{3/2}}
 \end{align}
 
 where $P_{\rm out}$ is the period of the outer binary and $m_{\rm tot}$ is the total mass of the triple system. 
 
 Due to large eccentricity of Pluto's orbit around the Sun, the effective quasi-secular corrections are encapsulated by $\tilde \epsilon_{\rm SA}=\epsilon_{\rm SA} (1+2\sqrt{2}e_{\rm out}/3)$. 
 For $\cos i_0 \sqrt{1-e^2}\lesssim 9\tilde \epsilon_{SA}/8$, the fluctuations in the angular momentum are larger than its initial value. The evolution experiences orbital flips, and the eccentricity is unbound, which hallmarks the transition to the non-secular regime, given by
 
 \begin{align}
\alpha_t = 3^{1/3} \left[\frac{128}{135} \frac{(1+e_{\rm{out}})^3}{\left(1+\frac{2\sqrt{2}}{3}e_{\rm{out}}\right)^2} \left(\frac{M_\odot}{m_{\rm{in}}}\right)^{1/3} \frac{R_{\rm{tot}}}{a_{\rm{out}}}\right]^{1/4}\approx 0.17
\end{align}

The regime that describes non-secular collisions is colored with blue -- non-secular collisions.

\section{Numerical results}\label{sec:numerical}

In order to verify the analytic result and simulate the dynamics of the progenitors of Pluto-Charon and the Sun, we used the publicly available N-body code \texttt{REBOUND} \citep{Rebound2012}. We use \texttt{IAS15}, a fast, adaptive, high-order integrator for gravitational dynamics, accurate to machine precision over a billion orbits \citep{Rebound_IAS}.
We integrate different sets of initial conditions; in all of them we set outer semi-major $a_{\rm{out}}=39.482 \rm{AU}$, outer eccentricity $e_{\rm{out}}=0.2488$ and changing mutual inclinations and outer semi-major axes. \cite{Canup2005} suggests a range of possible parameters for Pluto and Charon progenitors; we use the average masses from the constrained mass ranges, i.e. $m_{\rm{Pluto}}\approx 1.57\times 10^{24} \rm{g}$ and $m_{\rm{Charon}}\approx 1.35\times 10^{24} \rm{g} $. The total radius of the object is taken to be $R_{\rm{tot}}= R_{\rm{Pluto}}+R_{\rm{Charon}}\approx 1794 \rm{km}$ .

 \begin{figure*}
    \includegraphics[width=0.9\textwidth]{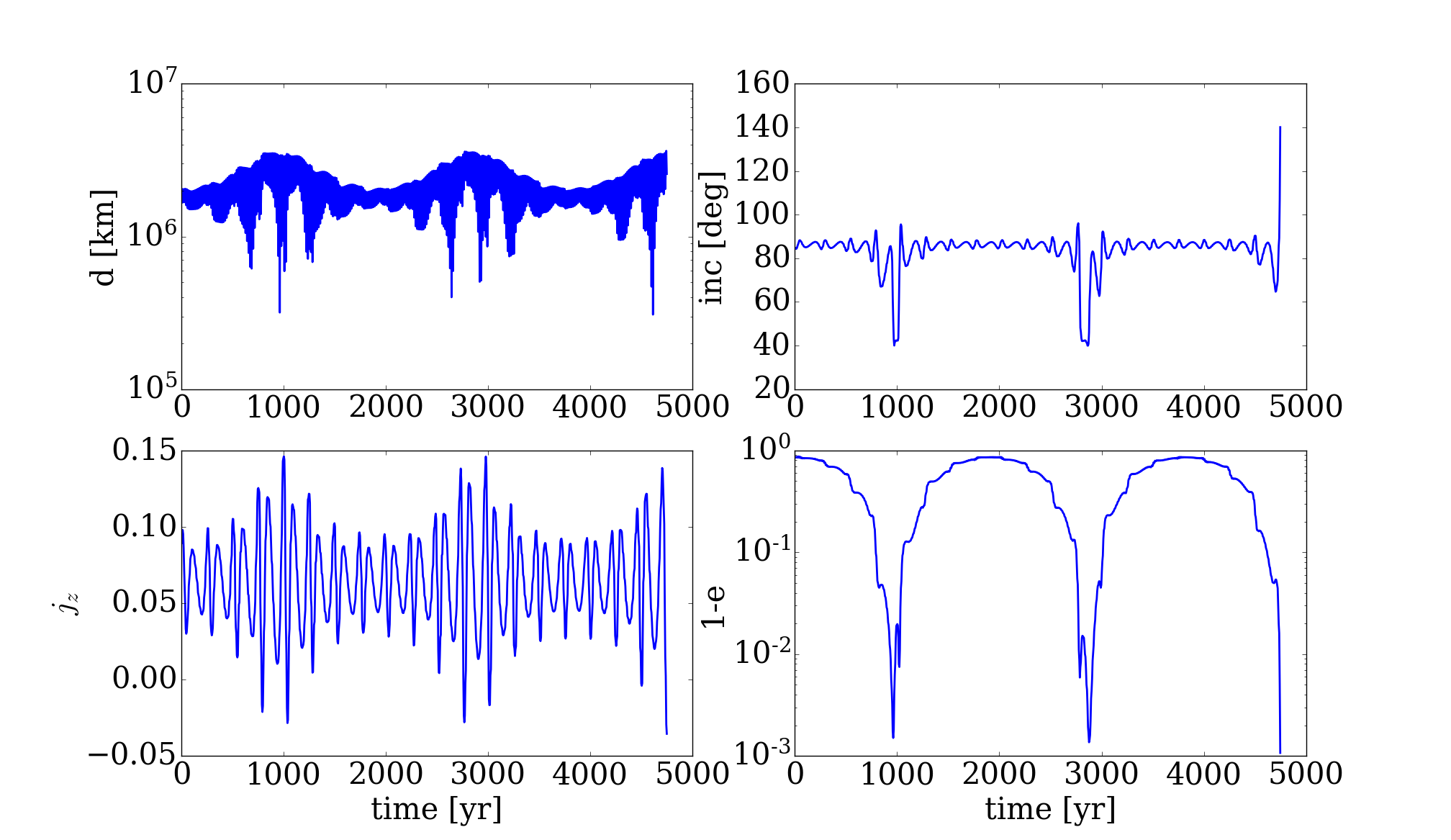}
\caption{ Evolution of the system for the following initial conditions: scaled initial inner semi-major axis $\alpha=0.3$, initial inner eccentricity $e_{\rm{in}}=0.15$, initial mutual inclination $i_{\rm{mutual}}=85^\circ$, argument of periapsis $\omega_{\rm{out}}=0$, longitude of the ascending node $\Omega_{\rm{out}}=\pi/4$, and mean anomalies $\mathcal M_1=0$ and $\mathcal M_2=-\pi/4$. 
} 
\label{fig:qsecular}
\end{figure*}

\begin{figure*}
    \includegraphics[width=\textwidth]{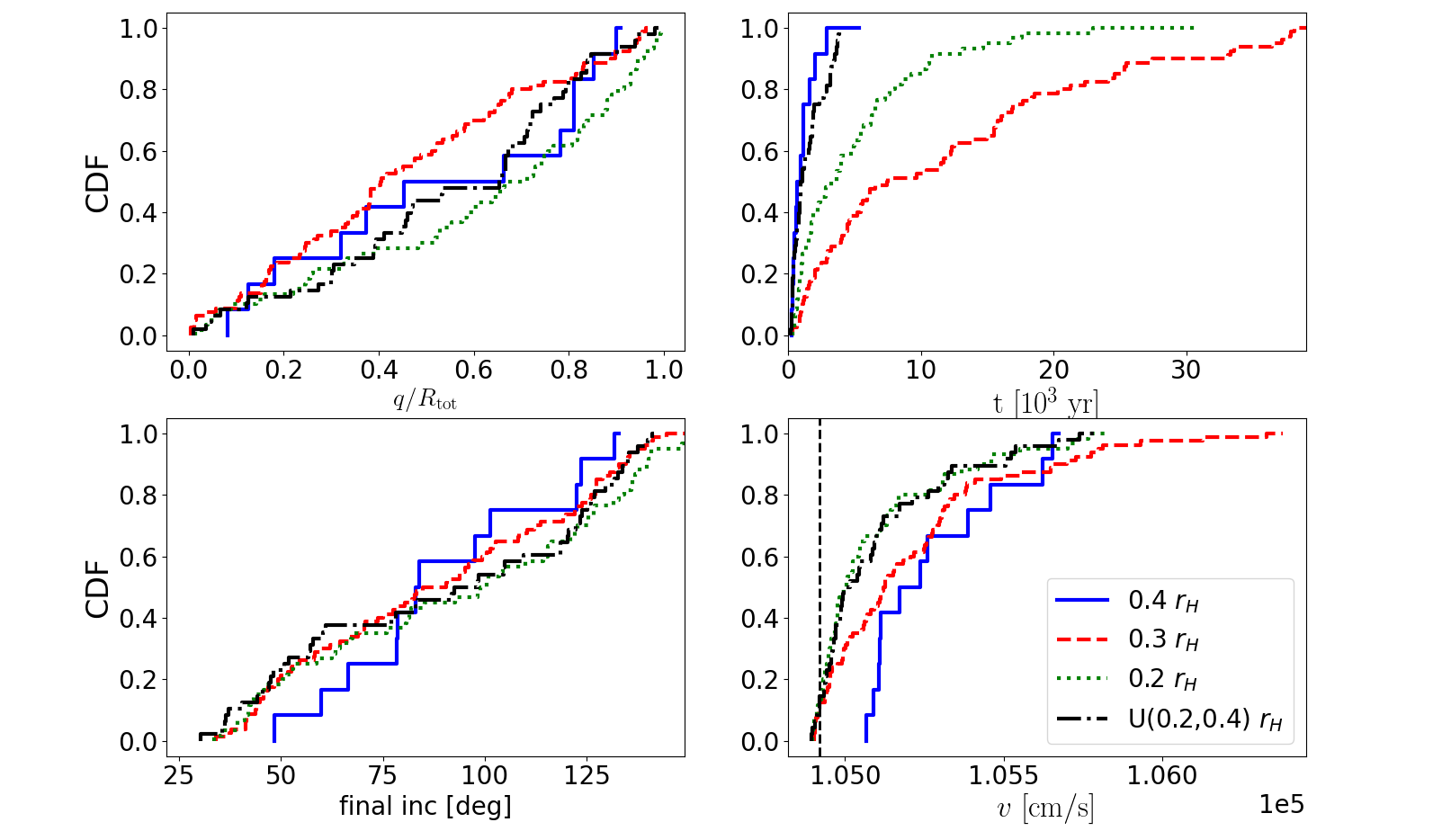}
\caption{ Cumulative distributions of the impact properties. Upper left: Normalized pericenter $q/R_{\rm{tot}}$; Upper right: Time of collision; Lower left: Final inclination at impact; Lower right: Velocity at impact. The vertical dashed line is the escape velocity. 
} 
\label{fig:CDF}
\end{figure*}

Fig. \ref{fig:qsecular} presents an example of the behavior in the quasi-secular regime. Here we study the evolution of the system with initial normalized inner separation of $\alpha=0.3$ in which case the evolution is in the non-secular regime. The inclination flips, the eccentricity is excited significantly during the evolution, and the inclination reaches a high value at the time of the collision. The angular-momentum in the $z$ direction, defined by $j_{\rm{z}}=\cos i \sqrt{1-e^2}$ oscillates, and its envelope structure can be derived analytically \citep{LuoKatz2016}. Collision occurs after $\sim 4500 \ \rm{years}$.  

In order to study the statistics of the collisional behavior of the Pluto-Charon binary system, we follow the approach of \cite{ArokothNature} and study a sample  of cases where $\alpha$ is either $0.2, 0.3, 0.4$, or chosen randomly from a uniform distribution in the range $[0.2,0.4]$ and run the simulation for $5\times 10^4 \ \rm{years}$. 
In Fig. \ref{fig:CDF} we show the cumulative distribution function of collision parameters.
 Fig. \ref{fig:CDF} present the CDF of the collision parameters: $q/R_{\rm{tot}}$ where $q=a(1-e)$ is the closest approach, time in units of thousand years, final inclination in degrees and velocity in units of $\rm{cm/sec}$. 
 
 The successful collisions fractions after $5\times 10^4 \rm \ years$ are: for $\alpha=0.2$ is $\approx 30\%$, for $\alpha=0.3$ is $\approx 40\%$, for $\alpha=0.4$ is $\approx 5\%$ and for the uniformly sampled $\alpha\in [0.2,0.4]$ is $\approx 20\%$. 
 
The lower left panel of Fig. \ref{fig:CDF} show the consistency with the uniform distribution in $\cos i_0$ where $i_0$ ranges between $40^\circ$ and $140^\circ$.
As can be seen from the upper right panel, the typical collision timescale is a few thousands of years and the collision velocity is close to the escape velocity -- lower right panel, which is given by $v_{\rm esc}=\sqrt{2G(m_{\rm Pluto}+m_{\rm Charon})/R_{\rm tot}}\approx 10.49\times 10^4 \rm{cm/sec}$, where $R_{\rm tot}$ is the minimal possible distance between the binary companions Pluto and Charon, i.e. the sum of their radii. 
As expected, since all the sampled values of $\alpha$ are predicted analytically to be in the regime of quasi-secular/non-secular or no collision at all, the behavior of the system is chaotic, and spans over wide range in the parameter space, as can be seen in the upper left panel.
Collisions in large values of $\alpha$, i.e. $\alpha \geq 0.4$, become more rare, since the system is less stable \citep{Gri+17}.

\section{Caveats}\label{sec:caveats}
In wider, and less hierarchical systems, the external perturbations by the Sun could lead to instabilities, and consequently lead to physical collisions or the escape of objects in the system through a chaotic evolution. While collisions are the major consequence discussed in this paper, escape is an unwanted byproduct.
Very wide systems can also become unstable due to the perturbations flyby encounters with other KBOs (Heggie's law \citealp{Heggie1975}), but here we neglect such encounters, and analyze only isolated systems.

 Analysis of the simulated systems show signatures of instability around $\alpha=0.4$, and as can be seen in Fig. \ref{fig:CDF}. Close to and beyond this limit the number of collisions decreases significantly. We find (see Fig. \ref{fig:CDF}) that $\approx 94\%$ of the systems with $\alpha=0.4$ break-up during the simulation.  For initially circular systems \cite{Gri+17} showed that the stability of systems could be sustained for even slightly wider systems; the somewhat lower limit we observed results from considering systems with significant eccentricities. 

Another caveat arises from uncertainties in the mass estimates, resulting from uncertainties in the chemical differentiation of the progenitors (see a detailed discussion in \citep{Stern2018_Pluto_after_NH}, and references therein) which we didn't take into consideration in our paper. 
The chosen masses are taken to be the average masses of the range given by \cite{Canup2005}, which might lead to some small uncertainties in our results. Furthermore, some larger uncertainties arise from the unknown initial separations between the progenitors of Pluto and Charon. We sampled some possible separations in order to explore a range of possibilities, but different initial separations could change the timescales and evolution of the binary.

We treated the Pluto-Charon - Sun system in isolation. In principle, other planets, in particular Neptune might affect the evolution of the system and add non-trivial corrections.

\section{summary}\label{sec:summary}
In this paper we proposed a novel formation channel for the origin of the Pluto-Charon system from a wide-binary, via secular and quasi-secular evolution that might lead to the collision between the components of the originally wide-binary progenitor. 

We use analytic criteria to set the different regimes of evolution: secular (and quasi-secular), non-secular and non-collisional, and made use of N-body simulations to verify and study these regimes. Our results indicate that collisions, consistent with the impact required to explain Pluto-Charon properties, are a natural byproduct of secular and quasi-secular evolution of wide-binary progenitors. The required impact parameters can be reproduced from a a wide range of initial conditions, suggesting this scenario as a robust formation channel for the origins of the Pluto-Charon system, and alleviating potential fine-tuned conditions required for the currently suggested origin from a random low-velocity collision between two of the most  massive, and relatively rare KBOs in the Kuiper-belt. The model could be used in the future for similar systems in the Solar-system and beyond, and shed light on the formation process of highly inclined contact or tidally synchronized binaries where regular LK oscillations that some of them could not have produced such configurations include (139775) 2001 QG$_{298}$ \citep{Lacerda2011} and
potentially 67P/Churyumov–Gerasimenko \citep{Marsden1969}, the latter one might be produced by a regular LK mechanism, since its obliquity is high but not extreme. Very large obliquities are a signature of quasi-secular evolution, but even large but not extreme obliquities might belong to the quasi-secular regime. 

\section*{Acknowledgements}
 
HBP acknowledges support from the MINERVA center for "Life under extreme planetary conditions" and the Kingsley fellowship at Caltech.

\section*{Data Availability}

The data that support the findings of this study are available from the
corresponding author upon reasonable request.



\bsp	
\label{lastpage}
\end{document}